\documentclass[prd, aps, nofootinbib, preprintnumbers, showpacs,
superscriptaddress, twocolumn]{revtex4}

\usepackage{graphicx,epsfig}

\usepackage{amssymb,amsmath}
\usepackage[english]{babel}
\usepackage{graphicx}
\usepackage[latin9]{inputenc}
\usepackage{epstopdf}
\usepackage{color}

\newcommand{\diag}{\mbox{diag}}
\newcommand{\diff}{d}

\newcommand{\p}{\partial}

\newcommand{\Sch}{\mbox{{\scriptsize Sch}}}
\newcommand{\Min}{\mbox{{\scriptsize Min}}}

\newcommand{\vp}{\phi}
\newcommand{\vt}{\theta}
\newcommand{\lm}{(\ell, m)}
\newcommand{\even}{\mbox{{\scriptsize even}}}
\newcommand{\odd}{\mbox{{\scriptsize odd}}}
\newcommand{\tot}{\mbox{{\scriptsize tot}}}

\newcommand{\gtil}{\tilde{g}}
\newcommand{\ghat}{\hat{g}}

\newcommand{\nabhat}{\hat{\nabla}}

\newcommand{\epshat}{\hat{\epsilon}}

\def\ii{{\rm i}}

\newcommand{\be}{\begin{equation}}
\newcommand{\ee}{\end{equation}}

\begin{document}

\title{How far away is far enough for extracting numerical waveforms,
  and how much do they depend on the extraction method?}

\author{Enrique Pazos}

\affiliation{Department of Physics and Astronomy, 202 Nicholson Hall,
  Louisiana State University, Baton Rouge, LA 70803, USA}

\affiliation{Center for Computation \& Technology, 216 Johnston Hall,
  Louisiana State University, Baton Rouge, LA 70803, USA}

\affiliation{Departamento de Matem\'atica, Universidad de San Carlos de Guatemala, 
 Edificio T4, Facultad de Ingenier\'{\i}a, Ciudad Universitaria Z. 12, Guatemala}

\author{Ernst Nils Dorband}

\affiliation{Department of Physics and Astronomy, 202 Nicholson Hall,
  Louisiana State University, Baton Rouge, LA 70803, USA}

\affiliation{Center for Computation \& Technology, 216 Johnston Hall,
  Louisiana State University, Baton Rouge, LA 70803, USA}

\author{Alessandro Nagar}

\affiliation{Dipartimento di Fisica, Politecnico di Torino,
  Corso Duca Degli Abruzzi 24, 10129 Torino, Italy}

\affiliation{INFN, sez.\ di Torino, Via P. Giuria 1, Torino, Italy}

\author{Carlos Palenzuela}

\affiliation{Department of Physics and Astronomy, 202 Nicholson Hall,
  Louisiana State University, Baton Rouge, LA 70803, USA}

\author{Erik Schnetter}

\affiliation{Center for Computation \& Technology, 216 Johnston Hall,
  Louisiana State University, Baton Rouge, LA 70803, USA}

\author{Manuel Tiglio}

\affiliation{Department of Physics and Astronomy, 202 Nicholson Hall,
  Louisiana State University, Baton Rouge, LA 70803, USA}

\affiliation{Center for Computation \& Technology, 216 Johnston Hall,
  Louisiana State University, Baton Rouge, LA 70803, USA}

\date{December 22, 2006}

\begin{abstract}
  We present a method for extracting gravitational waves from
  numerical spacetimes which generalizes and refines one of the
  standard methods  based on the Regge--Wheeler--Zerilli
  perturbation formalism.  At the analytical level, this
  generalization allows a much more general class
  of slicing conditions for the
  background geometry, and is thus not restricted to
  Schwarzschild--like coordinates.  At the numerical level, our
  approach uses high order multi-block methods, which improve both the
  accuracy of our simulations and of our extraction procedure.  In
  particular, the latter is simplified since there is no need for
  interpolation, and we can afford to extract accurate waves at large radii
  with only little additional computational effort.  We then present
  fully nonlinear three-dimensional numerical evolutions of a distorted 
  Schwarzschild black hole in Kerr--Schild coordinates with an odd
  parity perturbation and analyze the improvement we gain from our
  generalized wave extraction, comparing our new method to the
  standard one. We do so by comparing the extracted waves with 
  one-dimensional high
  resolution solutions of the corresponding generalized Regge--Wheeler
  equation.

  We find that, even with observers as far out as
  $R=80\,M$---which is larger than what is commonly used in
  state-of-the-art simulations---the
  assumption in the standard method that the background
  is close to having Schwarzschild-like coordinates increases the
  error in the extracted waveforms considerably.  Even for our
  coarsest resolutions, our new method decreases the error by between
  one and two orders of magnitudes.  Furthermore, we explicitly see that the
  errors in the extracted waveforms obtained by the standard method do
  not converge to zero with increasing resolution.  That is, these
  errors are dominated by the extraction method itself and not by the
  accuracy of our simulations.  We analyze in detail the quasinormal
  frequencies of the extracted waves, using both methods.

  In a general scenario, for example a collision of compact
  objects, there is no precise definition of
  gravitational radiation at a finite distance, and gravitational wave
  extraction methods at such distances are thus inherently approximate. 
  The results of this paper bring up the
  possibility that different choices in the wave extraction procedure
  at a fixed and finite distance may result in relative differences in the
  waveforms which are actually larger than the numerical
  errors in the solution. 
\end{abstract}


\maketitle

\section{Introduction}
One of the goals of numerical solutions of Einstein's equations is
usually the prediction and analysis of the gravitational radiation
emitted in some physical process.    There are many methods for computing, or
\emph{extracting}, gravitational waves from a numerical spacetime.
They can be broadly divided into two groups, depending on whether the
solution includes null infinity (or a portion of it), or whether the
computational domain is truncated at a hopefully large but finite
distance from the source.  In the first case, gravitational radiation
can be defined and extracted in an unambiguous, rigorous way (see
e.g.\ \cite{Winicour98} and references therein, and \cite{Huebner01}).
In the second case, some approximation has to be made; not only at the
level of the observer being in the radiation zone, but also in the way
the ``gravitational radiation'' is computed in terms of the spacetime
metric.  Due to the additional complexity of evolving Einstein's
equations all the way up to null infinity, currently most
simulations actually truncate the computational domain by
placing an artificial outer boundary at a finite distance.  This paper
deals with one particular approach to gravitational wave extraction
from spacetimes within this second group.

In general,  one expects the differences between the exact 
waveforms and those extracted at a finite distance 
 to decay as the extraction radius increases. 
One natural question that arises is: for a given extraction method,
how far away is
far enough,so that the errors in the
extracted waves are dominated by the accuracy of the simulations
used to obtain the numerical spacetime, and not by the extraction mechanism
itself?  In this paper we address this question in detail in a very
particular scenario, but which might shed some light on the general case. 

The main idea of extracting waves at a finite distance is to exploit
the structure of an asymptotically flat spacetime.  One reads off the
quantities which are needed to compute the gravitational radiation
from the numerically generated solution.  The method which we consider
in this paper is based on the well-known perturbations of the
Schwarzschild spacetime.
See e.g.\ \cite{Bruegmann:2006at, Buonanno06imr,
  Baker:2001sf, Zlochower2005:fourth-order, Fiske05, Friedrich96,
  Campanelli99,Beetle:2004wu, Nerozzi:2004wv, Burko:2005fa,
    Nerozzi:2005hz, Campanelli:2005ia} for other approaches based on the Weyl scalar
$\Psi_4$.

One
possible approach is to assume that the full metric in the region of
extraction can be considered as a perturbation of a flat spacetime,
and to read off such perturbations from the numerical solution.
This approach is justified by the fact that the leading order of the
metric at large distances (in an expansion in powers of $1/r$) is
flat.  If the waves are extracted at a large but finite distance from
the source, it makes sense to try to decrease the errors of the
approximation by further considering the next order in the expansion
of the metric, which is described by the Schwarzschild solution.  In
doing so, the numerical metric is not considered anymore a
perturbation of flat spacetime, but instead of the Schwarzschild
geometry.  
One can consider even higher orders in this
background identification, such as the spin contribution.  
However, an important fact to keep in mind
is that all these methods should in principle give the same gravitational radiation
as the radius of extraction increases.  In other words, one
should be able to compute the gravitational radiation through, for
example, perturbations of a flat spacetime or the Schwarzschild metric,
and the radiation 
should contain
the information about the spacetime's non-zero mass 
and---if present---angular momentum when the observer is at large 
enough distances. 

If only the first or the first two orders in the asymptotic expansion
of the metric are kept when identifying this distant ``background''
geometry, then the
framework for extracting gravitational radiation is that of
perturbations of flat spacetime or of the Schwarzschild geometry,
respectively.  One can view the former as a subcase of the latter, so
that from hereon we will just consider perturbations of the Schwarzschild
spacetime.  In this case, perturbations decouple into two separate
sectors, which
differ in the parity of the perturbations (odd or even).  These two
parity sectors are directly related to the real and imaginary parts of
the Weyl scalar $\Psi_4$ (see, for example,
ref.~\cite{Sarbach:2001qq}).  Gauge invariant formalisms for such
perturbations were developed by Regge
and Wheeler \cite{Regge57} in the fifties for the odd-parity sector and
by Zerilli \cite{Zerilli70} in the seventies for the even-parity
sector. 

The idea of using Regge--Wheeler--Zerilli perturbation theory to
extract gravitational waves from numerical spacetimes is definitely
not new.  It goes back to pioneering work by Abrahams and Evans
\cite{Abrahams88b, Abrahams89, Abrahams90} (see also \cite{Abrahams95b}) and it 
 has been used extensively since the birth of
numerical relativity  (see 
 \cite{Nagar:2005ea} for a review).  
For example, the accuracy of simulations of distorted black
holes was tested by comparing extracted waveforms against perturbative
calculations \cite{Camarda97b, Camarda97c, Brandt98, Alcubierre99b,
  Baker99a}, and often, also technical improvements (such as excision)
were tested by studying their effects on waveforms
\cite{Alcubierre01a, Alcubierre2003:BBH0-excision}.  Recently,
\cite{Herrmann:2006ks} reported Zerilli waveforms from
unequal mass binary black hole inspirals.  
In hydrodynamical simulations, gravitational waves are often
determined via the quadrupole formula, which usually gives more
accurate information in these particular situations (unless a black
hole is present), since the wave amplitude is typically very small and
thus difficult to detect from the spacetime metric \cite{Shibata02a,
  Shibata:2004kb, Duez04}.

In this paper we present a generalization of this approach to gravitational wave extraction
with two salient features.  The first is
at the level of the perturbation formalism itself: we use a
generalization of the standard Regge--Wheeler--Zerilli (RWZ)
formalism, which is not only gauge invariant, but also covariant
\cite{Gerlach79, Sarbach:2001qq, Martel:2005, Gundlach00b}, 
in the sense that it
is independent of the background coordinates.
The standard RWZ formalism is gauge
invariant only in the sense that the background metric is fixed to the
Schwarzschild geometry \emph{in Schwarzschild coordinates}, and the
formalism is invariant with respect to infinitesimal, first order
changes of coordinates, which keep the background coordinates
\emph{fixed}.  However, in numerical simulations of Einstein's
equations, the numerical spacetime might be close to the Schwarzschild
geometry in certain situations (say, at large distances), but the
metric does not need to be close to the Schwarzschild metric \emph{in
  Schwarzschild coordinates}.  In fact, when dealing with the black
hole singularity through black hole excision, one uses coordinates
that are well defined in a neighborhood of the horizon, and which are
therefore clearly not of Schwarzschild type.

This first salient improvement (the use of a generalized formalism)
is independent of the details of the numerical implementation.  The
second improvement is tied to our particular numerical approach, which
uses high order methods (typically higher than four) for high
accuracy, and uses multiple blocks with adapted grids, non-trivial
topologies, and smooth boundaries.  The use of high order methods for
both the evolution of Einstein's equations and for the wave extraction
procedure itself, combined with the use of shells of ``spherical''
patches or blocks, allows us to extract gravitational waves in a
simple, fast, and accurate way.  In particular, \emph{we can keep both
  the angular and radial resolutions fixed} and place the outer
boundaries at large distances, using considerably less computational
resources than what would be needed with Cartesian grids, even when
using mesh refinement.  In addition, no interpolation to spheres is
needed to extract waves on spherical shells.

For weak perturbations of a Schwarzschild black hole we can actually
obtain the exact solution by evolving the generalized
Regge--Wheeler equation.  Since this is a wave equation in $1+1$
dimensions, we can solve it with almost arbitrarily high accuracy.
For all practical purposes, we consider it to be an exact solution,
against which we can compare the
extracted waveforms from our three-dimensional evolutions.
We evolve weak perturbations of a
Schwarzschild black hole in Kerr--Schild coordinates, using the fully
nonlinear Einstein equations.
We find that the assumption in the standard method that the background
is in Schwarzschild--like coordinates increases the error in the
extracted waves (as compared to extracting with the correct
background) by between one and two orders of magnitudes.
This is true even with observers as far away as $80\,M$, and even for
the coarsest resolutions that we use.
Furthermore, we explicitly see that the errors in the standard method
do not converge to zero with increasing resolution at any fixed
extraction radius, while they do with the generalized method.
The errors only decrease (as $1/r$, as we discuss in sect.~\ref{simulations}) 
as the observer radius is increased.
That is, if one does not
use the correct background coordinates, these errors are dominated by
the extraction procedure and not by the accuracy of the simulations.
We compare the quasinormal frequencies of the waves extracted the
above methods against the results predicted by perturbation theory.

The organization of this paper is as follows.  In sect.~\ref{rw} we
describe in a self-contained way the generalized perturbation
formalism, restricted to the odd-parity sector (we will present a similar 
treatment for the even parity 
sector elsewhere), and our construction
of the Regge--Wheeler function from a numerical spacetime.  We also
use the inverse problem (that is, the
generation of a perturbed metric from any given Regge--Wheeler
function) to construct initial data that automatically satisfies the
Einstein constraints when linearized around the Schwarzschild
spacetime, which does not necessarily need to be given in
Schwarzschild coordinates.  This
is the data that we later evolve and use in our numerical tests.

In sect.~\ref{formulation} we briefly describe our numerical
techniques, out formulation of Einstein's equations, and our outer
boundary conditions.  Finally, we present our numerical results in
sect.~\ref{simulations}.
We first show that our extracted covariant and
gauge invariant Regge--Wheeler function coincides very well with the
expected one from perturbation theory (which we obtain by solving the
$1+1$ generalized Regge--Wheeler equation) when we use the generalized
formalism to identify the background correctly.  After that, we
compare our covariant and gauge invariant extracted waveforms with 
 those obtained by the
traditional approach, which assumes that the background is either the
Minkowski spacetime in Minkowski coordinates, or the Schwarzschild
spacetime in Schwarzschild coordinates. In sect.~\ref{sec:conclusions}
we discuss these results in the broader context of gravitational wave
extraction for generic spacetimes. 

For completeness, in Appendix~\ref{harmonic_decomposition} we describe
in detail our conventions for tensor spherical harmonics
decompositions.

\section{Odd-parity perturbations of Schwarzschild and wave extraction} \label{rw}
This section summarizes the
results of the generalized formalism relevant for this paper. We
closely follow
the notation and presentation of ref.~\cite{Sarbach:2001qq}.

\subsection{The background metric and tensor spherical decomposition of the perturbations}

The generalized formalism  assumes that the total metric can be written as
\be
g_{\mu \nu}^{\tot} = g_{\mu \nu} + \delta g_{\mu \nu} \label{tot_metric}
\ee
where $g_{\mu \nu}$ describes the Schwarzschild geometry and $\delta
g_{\mu \nu}$ is, in some sense, a ``small'' correction. Further, it is
assumed 
that the four-dimensional manifold can be decomposed as the product 
of a two-dimensional
manifold $\mathcal{M}$ parametrized with coordinates $x^a$ ($a=0,1$) and a unit 2-sphere $S^2$ with coordinates 
$x^A$ ($A=2,3$), such that the background Schwarzschild metric takes the form
\begin{eqnarray}
  \label{eq_metric}
  \diff s^2 & = & \gtil_{ab}(t,r)\,\diff x^a \diff x^b +
  f^2(t,r)\,\ghat_{AB}\,\diff x^A \diff x^B \,.
\end{eqnarray}
Capital Latin indices refer to angular coordinates $(\vt,\vp)$  on
$S^2$, while lower-case ones refer to the $(t,r)$ coordinates. 
Here $\ghat_{AB}$ is the standard metric on the unit sphere,
$\gtil_{ab}$ denotes the metric tensor on the manifold $\mathcal{M}$,
and $f^2$ is a positive function. 
 If one uses 
an areal radius coordinate, then $f=r$, but we do not make such an 
assumption. Actually, as we discuss below, the fact that
our formalism is general enough to allow for $f=f(t,r)$
 has practical advantages in the wave extraction procedure.
 For simplicity, the metric
on the unit 2-sphere $S^2$ is assumed to be in standard coordinates: 
$\ghat_{AB} = \diag(1, \sin^2\vt)$.
Summarizing, we are assuming that the background Schwarzschild
metric is given in a coordinate system in which there is no angular
shift, but there can be a radial shift.  Note that there is no
assumption about the shift in the perturbation.

From a numerical relativity point of view, 
it is usually convenient to deal with the variables that appear in the $3+1$
split of spacetime. To this end, we
follow the notation of ref.~\cite{Sarbach:2001qq}
 and explicitly expand the components of the background Schwarzschild metric as 
\begin{eqnarray}
  \label{coordinate_metric}
  \diff s^2 & = &
  (-\alpha^2 + \gamma^2 \beta^2) dt^2 + 2\gamma^2 \beta dt \, dr +
  \gamma^2 \, dr^2
  \\\nonumber
  & & {} + f^2(d\vt^2 + \sin^2\vt \, d\vp^2)
\end{eqnarray}
where $\alpha$ and $\beta \equiv \beta^r$ are the background lapse and radial
shift vector, respectively, and 
$\gamma^2 \equiv \gtil_{rr}$. 
Since the background is spherically symmetric, it is convenient to
expand the perturbations in spherical harmonics,
\begin{eqnarray}
  \delta g_{\mu \nu} & = & \sum_{\ell = 1}^{\infty}
  \sum_{m=-\ell}^{\ell} \delta g_{\mu \nu}^{(\ell,m)} \,.
\end{eqnarray}

In the odd-parity sector there is no
perturbation for $\ell=0$. The dipole term, $\ell=1$, corresponds to the linearization
of the Kerr metric using the angular momentum of the spacetime as a parameter. 
Thus, for gravitational wave extraction 
we only need to consider perturbations with $\ell \geq 2$.  
These quantities can be parametrizes according to
\begin{eqnarray}
  \label{eq:perturbADM}
  \delta\beta_A^{\lm} & = & b^{\lm} S_A^{\lm}
  \\\nonumber
  \delta g_{rA}^{\lm} & = & h_1^{\lm} S_A^{\lm}
  \\\nonumber
  \delta g_{AB}^{\lm} & = & h_2^{\lm} S_{AB}^{\lm}
  \\\nonumber
  \delta K_{rA}^{\lm} & = & \pi_1^{\lm} S_A^{\lm}
  \\\nonumber
  \delta K_{AB}^{\lm} & = & \pi_2^{\lm} S_{AB}^{\lm} \, .
\end{eqnarray}
Using 
the covariant derivative $\nabhat_A$ compatible with the metric
$\ghat_{AB}$ on the unit sphere $S^2$ and its associated Levi--Civita
tensor $\epshat_{AB}$ (with non-vanishing components $\epshat_{\vt \vp}=\sin \vt =
-\epshat_{\vp \vt}$), we define 
$S_A = \epshat^B_{\; A} \nabhat_B Y$ (the first index in $\epshat$
raised with the inverse of $\ghat$) and
$S_{AB}=\nabhat_{(A} S_{B)}$. 
Here, $Y\equiv Y^{(\ell, m)}$ are the standard spherical harmonics.
The quantities $S_A$ and  $S_{AB}$ form a 
basis on $S^2$ for odd-parity vector and symmetric tensor fields,
respectively.
For completeness, 
we give a detailed and self-consistent
description of how to use these to decompose vectors and tensors into
spherical harmonics in Appendix~\ref{harmonic_decomposition}.

From now on, we suppress the superindices $(\ell, m)$ and the sum
over them, since  modes
belonging to different pairs of $(\ell, m)$ decouple from each
other  in the perturbation formalism.

\subsection{Extraction of the Regge--Wheeler function from a given geometry}
\label{sec:extract}

To define the background metric, we extract the $\ell=0$ component
(that is, the spherically symmetric part) of 
 the numerical solution $g_{\mu \nu}^{\tot}$.
This is done by decomposing the metric 
$\tilde{g}_{ab}$ of the
two-dimensional manifold $\mathcal{M}$ into spherical harmonics.
These metric components behave like scalars under a rotation of
coordinates. Thus, the background metric is computed as
\begin{equation}
  \tilde{g}_{ab} = \frac{1}{4\pi} \int g_{ab}^{\tot} d \Omega \,,
\end{equation}
where $d\Omega$ is the standard area element on $S^2$.  The
function $f$ can be computed  through $f=\sqrt{A/4\pi}$, with 
\begin{equation}
  A=\int \sqrt{\ghat} \, d\theta \, d\phi \,,
\end{equation}
where the integration is performed over the extraction 2-sphere, and
$\ghat$ is the determinant of $\ghat_{AB}$. 

Similarly, we compute the perturbed quantities by extracting the $\ell \ge
2$ components of the numerical metric $g_{\mu \nu}^{\tot}$, in the way 
explained in Appendix~\ref{harmonic_decomposition}.  

Once we have obtained the multipoles  
$b, h_1, h_2, \pi_1, \pi_2$ defined above in eq.~(\ref{eq:perturbADM})
and the background quantities $f$, $\alpha$, $\gamma$, $\beta$ defined
in eq.~(\ref{coordinate_metric}),
we can find the generalized gauge-invariant Regge--Wheeler (RW) function
$\Phi_{RW}$. It is given by \cite{Sarbach:2001qq}
\begin{equation}
\Phi_{RW}= \frac{2 f}{\lambda\alpha\gamma} \left(\alpha\,\pi_1 - \frac{\p_0
f}{f} h_1 \right) \label{pot_adm}
\end{equation}
where $\partial_0 \equiv \partial_t - \beta \partial_r$ and
$\lambda=(\ell-1)(\ell+2)$. 
Notice from eq.~(\ref{pot_adm}) that
the \emph{only} multipole components  
appearing in the RW function $\Phi_{RW}$ are $h_1$ and $\pi_1$, so
that there is no need to compute the others.

Previous approaches to compute waveforms 
with the standard RWZ formalism have typically been considerably more
involved than what we have just described.  We briefly sketch the standard
approach here.  Einstein's equations are usually solved using Cartesian 
coordinates on a Cartesian grid.  The numerically obtained metric is
first transformed to polar-spherical coordinates.
Performing the multipole decomposition on a given coordinate
sphere requires a numerical integration over that sphere,
which in turn requires interpolating the metric to the spherical
surface, which does not coincide with the grid points of the Cartesian
grid.
Integrating over the sphere also allows
computing the areal radius and its radial derivatives.
These quantities are then used to transform the metric in a second
step to its final form in
``Schwarzchild-like'' coordinates.  This is done by first changing from
the coordinate radius to an areal radius (which requires the numerically
calculated radial derivatives), and then identifying the
$(t,r)$ components of the metric in this new coordinate system,
which is assumed to be a perturbation of the Schwarzschild metric in
Schwarzschild coordinates.  With all this in
place, the waveforms are then computed using standard RWZ formulae. 

In our case, the multi-block grid structure naturally allows 
for spherical surfaces. Hence, no interpolation is required.
The generalized perturbation formalism allows us to compute the RW
function $\Phi_{RW}$
without transforming the metric to Schwarzchild coordinates.  In
particular, the transformation to an areal radial coordinate is not required
at all.  Thus, our
extraction procedure amounts simply to numerical integrations at a given
value of the radial coordinate to compute the
multipoles, and then using eq.~(\ref{pot_adm}) to compute the RW function.
An additional improvement is that our high order accurate derivative
operators are naturally associated with a high order accurate discrete
norm, leading to an integration procedure which has the same accuracy
as our derivative operators.

\subsection{(Re)construction of the metric from the Regge--Wheeler function} \label{reconstruction}

It can be seen (see, for example,
ref.~\cite{Sarbach:2001qq} for more details of what follows)
that for any slicing of Schwarzschild of the type given in
eqs.~(\ref{eq_metric}) or~(\ref{coordinate_metric}),
that we can construct a perturbed four-metric
from the RW potential.  The perturbation coefficients of the
linearized metric, as defined in eq.~(\ref{tot_metric}), become
\begin{widetext}
\begin{eqnarray}
\delta g_{r\vt } & = & \left[\frac{\gamma }{\alpha }\left( -f\dot{\Phi}_{RW} +
\beta f\Phi_{RW}' + \Phi_{RW}(\beta f' - \dot{f}) \right) + \frac{fk'-2kf'}{f}\right]
\frac{Y_{\vp}}{\sin{\vt }}  \\
\delta g_{r\vp } & = & -\left[\frac{\gamma }{\alpha }\left( -f\dot{\Phi}_{RW} + \beta f\Phi_{RW}' +
\Phi_{RW}(\beta f' - \dot{f}) \right) + \frac{fk'-2kf'}{f}\right]
 \sin{\vt }
Y_{\vt}  \\
\delta g_{\vt \vt } & = & \frac{2k}{\sin^2{\vt }}\left[-\cos{\vt
}Y_{\vp}+\sin{\vt }Y_{\vt \vp } \right] \\
\delta g_{\vt \vp } & = & k\left[\cos{\vt }Y_{\vt }+\sin^{-1}{\vt
}Y_{\vp \vp } -\sin{\vt}Y_{\vt \vt} \right] \\
\delta g_{\vt t } & = & \left[\frac{1}{\gamma \alpha }\left( -\gamma^2\beta f\dot{\Phi}_{RW}
 + f(\gamma^2\beta^2-\alpha^2)\Phi_{RW}' +
 (-\alpha^2f'-\dot{f}\beta \gamma^2+f'\gamma^2\beta^2 )\Phi_{RW} \right)
 + \frac{f\dot{k}-2k\dot{f}}{f} \right]\frac{Y_{\vp}}{\sin{\vt }}  \\
\delta g_{\vp \vp } & = & 2k \left[\cos{\vt
}Y_{\vp} - \sin{\vt }Y_{\vt \vp } \right] \\
\delta g_{\vp t } & = & -\left[\frac{1}{\gamma \alpha }\left( -\gamma^2\beta f\dot{\Phi}_{RW}
 + f(\gamma^2\beta^2-\alpha^2)\Phi_{RW}' +
 (-\alpha^2f'-\dot{f}\beta \gamma^2+f'\gamma^2\beta^2 )\Phi_{RW} \right)
 + \frac{f\dot{k}-2k\dot{f}}{f} \right]\sin{\vt }Y_{\vt }
\end{eqnarray}
\end{widetext}
Here dots and primes denote derivatives with respect to time and radius,
respectively.
It is $Y_{\vp} = \partial_{\vp}Y$, $Y_{\vt} = \partial_{\vt}Y$, and as before
we are skipping the $\lm$ superindices.  $\gamma$, $\alpha$, $\beta$,
and $f$
are defined in eq.~(\ref{coordinate_metric}). 
It can be seen that the function $k$ is a pure gauge term and completely arbitrary; in
particular, we can make it vanish (resulting in 
the so called Regge--Wheeler gauge) through a first order coordinate
transformation.

The generalized RW equation is 
\begin{widetext}
\begin{eqnarray}
\ddot{\Phi}_{RW} & = & c_1\dot{\Phi'}_{RW}+  c_2 \Phi^{''}_{RW} +
c_3\dot{\Phi}_{RW} + c_4\Phi'_{RW}   - \alpha^2 V \Phi_{RW}
\label{master}
\end{eqnarray}
with the coefficients
$c_i$ and the potential $V$ given by
\begin{eqnarray}
c_1 & = &  2 \beta  \\
c_2 &=&
 {\frac {\left (\alpha^2 - \gamma^2\beta^2 \right
)}{\gamma^2}}  \\
c_3 & = & {\frac {\left (\gamma  \dot{\alpha} - \gamma  \beta
\alpha ' +\alpha \beta \gamma ' - \alpha \dot{\gamma } + \gamma  \alpha \beta '
\right )}{\gamma  \alpha  }} \\
c_4 &=&
{\frac{1}{\gamma^3\alpha } \left (- \gamma^3\beta \dot{\alpha } - \alpha^3
\gamma ' + \gamma^3 \beta^2\alpha ' - 2 \gamma^3\alpha \beta \beta ' +
\gamma^3\alpha \dot{\beta } + \gamma^2\alpha \beta \dot{\gamma } + \gamma
\alpha^2\alpha ' - \gamma^2 \alpha \beta^2\gamma ' \right ) }
\\
V & = & \frac{1}{f^2}\left[ \ell(\ell+1) - \frac{6M}{f} \right] \,.
\end{eqnarray}
\end{widetext}
When the background metric is Schwarzschild in Schwarzschild
coordinates, this generalized RW equation coincides of course with
the standard equation.
 Below, in sect.~\ref{simulations}, we use high-resolution solutions of
this generalized $1+1$ equation as ``exact'' solutions ,
against which we compare the extracted RW function from our
three-dimensional distorted black 
hole simulations.

\section{Formulation of the equations,
  boundary conditions, initial data, and numerical methods}
\label{formulation}

\subsection{Evolution equations}

The numerical simulations shown below were performed by evolving a first
order symmetric hyperbolic reduction of the Generalized Harmonic 
formalism, as constructed in ref.~\cite{Lindblom:2005qh}.  In this formulation, the coordinates $x^\mu$
are chosen to satisfy the (generalized) harmonic condition%
\footnote{In this subsection we use the Latin indices 
  $i,j,k,\ldots$ to denote three-dimensional spatial quantities, while
  Greek indices 
  continue to represent the four-dimensional ones.}
\cite{Friedrich85}
\begin{eqnarray}
  \label{eq:harmonic1}
  \nabla^\sigma \nabla_\sigma x^{(\mu)} & = & H^{(\mu)}(t,x^i) \, ,
\end{eqnarray}
where the gauge source functions $H^{(\mu)}(t,x^i)$ are freely
specifiable functions of
the spacetime coordinates, and $\nabla_\mu$ is the covariant
derivative associated
with $g_{\mu \nu}$. Here we omit the label ``tot'' from the metric
(\ref{tot_metric}) for the sake of simplicity.
The reduction from second to first order is achieved by introducing the first
derivatives of the metric $g_{\mu \nu}$ as independent quantities.
Following ref.~\cite{Lindblom:2005qh}, we introduce the quantities 
\begin{eqnarray}
  \label{eq:first_order_quantities_Q}
  Q_{\mu \nu} & = & -n^\sigma \partial_\sigma g_{\mu \nu}
  \\
  \label{eq:first_order_quantities_D}  
  D_{i \mu \nu} & = & \partial_i g_{\mu \nu} \, ,
\end{eqnarray}
where $n_\mu=-\alpha \nabla_\mu t$ is the (future
directed) timelike unit normal vector to the hypersurface $t = x^0 =
\mathrm{const}$. Thus, the evolution equations for $Q_{\mu \nu}$ are given
by the Generalized Harmonic formalism, while the evolution equations for $D_{i \mu \nu}$
are obtained by applying a time derivative to their definition
(\ref{eq:first_order_quantities_D}) and commuting the temporal and
spatial derivatives.  Finally, the metric $g_{\mu \nu}$ is evolved  using
the definition of $Q_{\mu \nu}$, eq.~(\ref{eq:first_order_quantities_Q}).  In
addition, in the spirit of refs.~\cite{Brodbeck98,
  Gundlach2005:constraint-damping},
the constraints of this system are added to the
evolution equations in such a way that the physical solutions (i.e.,
those satisfying the constraints) are an attractor in certain
spacetimes.
In those situations, small constraint violations will be
damped during the evolution.  The whole construction of this
formulation of the equations is described in
detail in \cite{Lindblom:2005qh}.

The standard $3+1$ components of the metric (i.e., the lapse function
$\alpha$, the shift $\beta^i$, and the intrinsic metric $\gamma_{ij}$)
can be obtained via the relations
\begin{eqnarray}
  \label{eq:3+1fields}
  \alpha^2 & = & -1/g^{tt}
  \\
  \beta_i & = & g_{ti}
  \\
  \gamma_{ij} & = & g_{ij} \, .
\end{eqnarray}
The extrinsic curvature is defined in terms of the intrinsic
metric as
\begin{eqnarray}
  \label{eq:def_Kij}
  K_{ij} & = & \frac{1}{2 \alpha} (\partial_t - {\cal L}_{\beta})
  \gamma_{ij} \, .
\end{eqnarray}   
It can be recovered from the fields $Q_{\mu \nu}$, $D_{i \mu \nu}$ via
\begin{eqnarray}
  -2 \alpha K_{ij} & = & \alpha Q_{ij} + D_{itj} + D_{jti}
  \\\nonumber
  & & {} - \gamma^{km} \beta_m (D_{ijk} + D_{jik}) \, .
\end{eqnarray}
In our simulations below we also monitor the Hamiltonian and momentum
ADM constraints, namely,
\begin{eqnarray}
  \mathcal{H}   &=& \frac{1}{2} \left( {}^{(3)}R - K_{ij} K_{ij} + (tr K)^2 \right) \\
  \mathcal{M}_i &=& \nabla_k \left( K^k_i - \delta^k_i tr K \right)  \,,
\end{eqnarray}
where ${}^{(3)}R$ is the Ricci scalar associated to the three-dimensional
space-like metric $\gamma_{ij}$.

We impose maximally dissipative boundary conditions at the outer
boundary.  While these conditions guarantee well-posedness of the associated
initial value problem, and thus numerical stability with our particular
discretization, they are physically incorrect in the sense that they
do  not include
back-scattered radiation from outside the simulation domain.  For that
reason, in the simulations shown below we place the outer 
boundary at large enough distances so
that our extracted waves are causally disconnected from boundary effects.

\subsection{Multi-block approach}

We use multi-block (also called multi-patch or multi-domain) methods
for our numerical calculations.  These have several advantages over
single-domain methods:

\paragraph*{Smooth boundaries.}  They provide smooth outer and inner
boundaries, which is in general required \cite{Secchi1996a} for a
well-posed initial boundary value problem.

\paragraph*{Constant resolution.}  They allow us to use constant
radial and angular resolutions.  This is not possible with mesh
refinement methods.  The way in which mesh refinement is typically
used leads to a decreasing radial resolution, which makes it
difficult to extract accurate gravitational wave information in
the wave zone of a binary black hole system.  (See
sect.~\ref{sec:conclusions}, where we list typical wave extraction
radii and resolutions.)

\paragraph*{No time-stepping restrictions.}  They do not lead to a
deterioration of the CFL factor for co-rotating coordinates.  (For
example, \cite{Bruegmann:2003aw} reports that the CFL factor had to be
reduced on the outermost refinement levels.  The same was done later
in \cite{Diener-etal-2006a}.)

\paragraph*{Adapted to symmetries.}  They can be adapted to the
symmetries of a system.  Obviously, adapted coordinates can reduce the
discretization error significantly.  In our case, we use on each block
one radial and two angular coordinates to model the geometry of a
single black hole.  For binary black hole systems, one can use blocks
that are roughly spherical near the individual holes and far away in
the wave zone, with a transition region in between.  Fig.~5 in
\cite{Schnetter06a} shows a possible multi-block system for this.

\paragraph*{No coordinate singularities.}  They have no coordinate
singularities.  Spherical or cylindrical coordinates have
singularities on the $z$ axis which may cause problems.  An
alternative approach which avoids these singularities would be to use
a pseudo-spectral decomposition into spherical harmonics.
This was used in
\cite{Scheel-etal-2003:scalar-fields-on-Kerr-background} to evolve
scalar fields on a Kerr background, in \cite{Kidder:2004rw,
  Scheel-etal-2006:dual-frame} to evolve Einstein's, and in
\cite{Gourgoulhon01, Grandclement02, Pfeiffer:2002wt, Ansorg:2004ds,
  Ansorg:2005bp, Loeffler06a} to set up initial data for various black
hole and neutron star configurations.

\begin{figure}
  \includegraphics[width=0.45\textwidth]{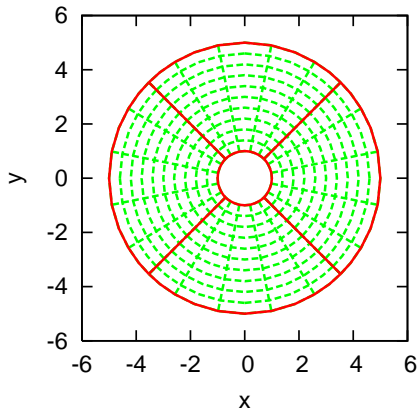}
  \caption{\label{fig:six-patches}%
    The equatorial plane of an example six-block geometry, cutting
    through four blocks.  Note that the blocks do not overlap.  All
    six blocks are made up identically.  The outer and inner
    boundaries are smooth spheres.  The outer boundary in our 
    typical simulations is actually located much further out than shown here.}
\end{figure}

Of course, using multiple blocks adds to the complexity of an
implementation.  However, the properties of multi-block
systems for hyperbolic equations are by now well understood, and we describe
our particular approach in \cite{Lehner2005a} and \cite{Diener05b},
and in some detail in \cite{Schnetter06a}.  In this particular paper we use 
a \emph{six-block system} to discretize the geometry of a single black
hole, which is depicted in fig.~\ref{fig:six-patches}.  We use the
same tensor basis on each block, namely a global three-dimensional
Cartesian coordinate system.  We found that this greatly simplifies
the inter-block boundary conditions, since all components of tensorial
quantities are then scalars with respect to the block-local coordinate
systems.

We use the \emph{penalty method} to enforce the inter-block boundary
conditions.  The penalty method for finite differences is described in
\cite{Carpenter1994a, Carpenter1999a, Nordstrom2001a}, and we describe
our approach and notation in \cite{Lehner2005a, Diener05b,
  Schnetter06a}.  In short, the penalty
method works as follows.  The individual blocks do not overlap, but
they have their boundary points in common.  The evolution equations
are first discretized on each block independently using one-sided
derivatives near the block boundaries.  Then a correction term is
added to the right hand side of the time derivative of each 
characteristic variable at the boundary points, penalising the
difference between the left and right eigenmode values $u^l$ and
$u^r$ on the boundary points:
\begin{eqnarray}
  \partial_t u^l & \rightarrow & \partial_t u^l + \frac{
    S^l}{h^l \sigma^l}\, (u^r - u^l)
  \\
  \partial_t u^r & \rightarrow & \partial_t u^r + \frac{
    S^r}{h^r \sigma^r}\, (u^l - u^r) \,.
\end{eqnarray}
Here $h^l$ and $h^r$ are the grid spacings on the two blocks, which
may be different.  These penalty terms ensure continuity between
the two blocks in the continuum limit and numerical stability in the
semidiscrete case if the relevant parameters are appropriately chosen.
The quantities $\sigma^l$ and $\sigma^r$ depend on the coefficients of
the differencing operators that are used on the two blocks.%
\footnote{To be exact, $\sigma^l$ and $\sigma^r$ depend on the
  coefficients of the discrete norms that are used in the blocks, but
  the differencing operators and the norms are usually chosen together
  to satisfy \emph{summation by parts}.  See \cite{Calabrese:2003a,
    Calabrese:2003vx,
    Lehner-etal-2004:Kerr-cubical-excision-problems}, and especially
  \cite{Diener05b}.}
The parameters $S^l$ and $S^r$ determine how much (if any) dissipation
is introduced across the block boundary.  To ensure stability, they must be
chosen in a very specific way depending on the characteristic
speeds of the evolution system.

We have implemented this in the Cactus framework \cite{Goodale02a,
  cactusweb1}, using the Carpet driver \cite{Schnetter-etal-03b,
  carpetweb} and the CactusEinstein toolkit \cite{cactuseinsteinweb}.

\subsection{Initial data}

If the RW function $\Phi_{RW}$ satisfies the RW equation
(\ref{master}), then the
perturbed metric constructed in sect.~\ref{reconstruction} satisfies the
linearized Einstein equations.  Furthermore, it can be explicitly
shown that this metric initially 
satisfies the linearized constraints around the Schwarzschild geometry
for \emph{any} initial values $\Phi_{RW}(t=0,r)$ and
$\dot{\Phi}_{RW}(t=0,r)$.%
\footnote{When constructing initial data for the $3+1$
  quantities, one also needs to take time derivatives of the
  four-metric; where second time derivatives of $\Phi_{RW}$ appear, we use
  the RW equation to trade these for space derivatives.}  We take
advantage of this property and construct initial data in a simple way
as a test our new wave extraction method.  For our simulations below, we
use Kerr--Schild coordinates for the Schwarzschild background, and for
the distortion we set $\ell=2$, $m=0$, and choose
\begin{eqnarray}
  \label{eq:InitialData}
  \Phi_{RW}(t=0,r) & = & 0 \, ,
  \\\nonumber
  \dot\Phi_{RW}(t=0,r) & = & A\, e^{(r-r_0)^2/\sigma^2}
\end{eqnarray}
with parameters $r_0$ and $\sigma$.  This corresponds to a Gaussian
pulse of width $\sigma$ centered at $r=r_0$.
 
If we assume that we can Taylor--expand
(a suitable norm of) the discrete non-linear constraints in terms of
the perturbation amplitude $A$ for any fixed gridspacing $h$, we have
\begin{eqnarray}
  \label{eq:IDConstraintsTaylor}
  \mathcal{C}(A,h) & = &
  \left. \mathcal{C}(A,h) \right|_{A=0}
  \\\nonumber
  & & {} + A \left. \frac{\partial \mathcal{C}(A,h)}{\partial A}
  \right |_{A=0}
  \\\nonumber
  & & {} + \frac{A^2}{2} \left. \frac{\partial^2
      \mathcal{C}(A,h)}{\partial A^2} \right |_{A=0}
  + \mathcal{O}(A^3) \, .
\end{eqnarray}
Since in the continuum the linearized constraints are satisfied, the
first two terms in the above expansion 
vanish for
$h\rightarrow 0$, but otherwise are of the order of the truncation
error. For small enough  $A$ the first term (that is, the background 
contribution) dominates, and the term $C(A,h)$ appears to be independent of
$A$.  For large enough $A$, on the other hand, 
the quadratic term in the expansion given by eq.~(\ref{eq:IDConstraintsTaylor})
will dominate.
 
\begin{figure}
  \begin{center}
    \includegraphics[width=0.45\textwidth]{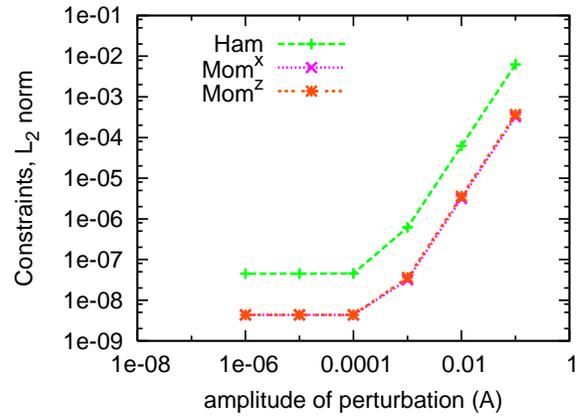}
    \caption{Discrete constraint violations for various perturbation
      amplitudes $A$
      at a fixed (high) resolution.  We show the $L_2$ norm for
      the Hamiltonian constraint and for two components ($x$ and $z$) of the
      momentum constraint (which turn out to be very close to each other, as
      the plot shows).   
      The numerical resolution is $109\times109$ grid points per block in the angular
      directions and $\Delta r \approx 0.0148$ in the radial
      direction.  The behavior is as
      expected and as described in the body of the paper: for
      sufficiently small
      amplitudes, the background contribution dominates the
      discretization error in the constraints, which then appear to
      be independent of $A$.  For large enough amplitudes,
      the constraint violation has a quadratic dependence on $A$ (with
      an exponent of $2.01 \pm 0.01$ for the resolution
      shown in this figure), since for our initial data only the
      linearized constraints (around
      Schwarzschild) are satisfied.}
    \label{fig:IDConstraintsScaling}
  \end{center}
\end{figure}

\begin{figure}
  \begin{center}
    \includegraphics[width=0.45\textwidth]{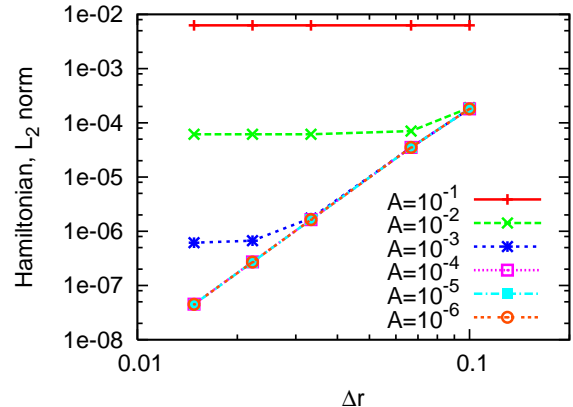}
     \caption{$L_2$ norm of the Hamiltonian constraint for different
      amplitudes $A$ of the perturbation and for different
      resolutions $h$.  The coarsest resolution uses
      $17\times17$ points per block in the angular directions and
      $\Delta r = 0.1\,M$ in the radial direction.  We increase the
      resolution in
      all directions, up to $109\times109$ points in the angular
      directions and
      $\Delta r \approx 0.0148\,M$ in the radial direction.
      Since only the linearized
      constraints are satisfied, the non-linear constraints do not
      converge to zero.  For sufficiently large perturbation amplitudes and
      for sufficiently fine resolutions, the non-linear effects become
      visible, and the constraint violations converge to a constant
      value which depends on the amplitude $A$.  As shown in the
      previous figure, this dependence is quadratic, as
      expected. }
    \label{fig:IDConstraints}
  \end{center}
\end{figure}

Fig.~\ref{fig:IDConstraintsScaling} presents numerical evidence
that this expected behavior is indeed the case.  We set up
numerical data according to eq.~(\ref{eq:InitialData}),
with perturbation amplitudes $A$ between
$10^{-6}$ and $10^{-1}$.  The radial domain extent is $1.8 \le r \le
7.8$, the perturbation is centered around $r_0=4.8\,M$ and has a width
of $\sigma=1.0\,M$.  We then compute the
discrete Hamiltonian and momentum constraints
$\mathcal{H}$ and $\mathcal{M}^i$ for these initial data sets,
using the same (high) 
resolution, namely $109\times 109$ grid points on each block in the
angular direction and $406$ points in the radial direction,
corresponding to $\Delta r \approx 0.0148\,M$.  Due to the symmetry
of our six-block structure and the axisymmetry of the initial data, 
two components of the discrete momentum constraints 
coincide, $\mathcal{M}^x = \mathcal{M}^y$, and we therefore do not
show the latter. 
The behavior of the constraints in the
$L_2$ and the $L_\infty$ (not shown in the figure) norms 
agrees with eq.~(\ref{eq:IDConstraintsTaylor}):
for small amplitudes $A$, the discrete
constraints at a fixed resolution appear to be independent of $A$, 
while for large amplitudes they show the expected quadratic
dependence on $A$. 
We also show that the discrete constraint violations of our initial
data sets 
have the expected dependence on resolution.
For small amplitudes and coarse resolutions,
the contribution of the quadratic term in eq.~(\ref{eq:IDConstraintsTaylor}) 
is sufficiently small, so that the
constraints seem to converge towards zero.  However, for any given
amplitude $A$ a fine enough resolution $h$ reveals
that the convergence is actually towards a
small but non-zero value, determined by the quadratic term in the
expansion eq.~(\ref{eq:IDConstraintsTaylor}).  This behavior is shown in
fig.~\ref{fig:IDConstraints}.  As an illustration we show there a
convergence test for $\mathcal{H}$ by comparing initial data for
different resolutions.  The highest resolutions are identical to those
used in fig.~\ref{fig:IDConstraintsScaling}.  The other four
resolutions shown are $73\times73\times271$, $49\times49\times181$,
$25\times25\times91$, and $17\times17\times61$ grid points per block,
corresponding to $\Delta r \approx 0.0222\,M$, $0.0333\,M$,
$0.0667\,M$, and $0.1\,M$, respectively.

\section{Numerical studies} \label{simulations}

\subsection{Description of the simulations}

We use the $D_{8-4}$ operator constructed 
in \cite{Diener05b}, a summation by parts operator
\cite{Kreiss1974a, Kreiss1977a}
which is
eighth order accurate in the interior and fourth order accurate at
the boundaries, optimized to minimize its spectral radius
and boundary trunctation errors.  Fifth order global convergence is expected 
\cite{Gustafsson1971:hyperbolic-BC-FD-convergence,
  Gustafsson1975:hyperbolic-BC-FD-convergence}. 
We 
integrate in time with a fourth order Runge--Kutta
integrator with adaptive time stepping as described in \cite{Press86}.

\begin{figure}
\begin{center}
\includegraphics[width=0.45\textwidth]{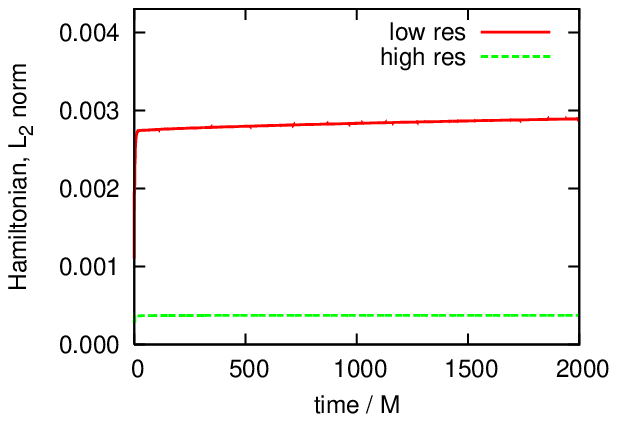}
\includegraphics[width=0.45\textwidth]{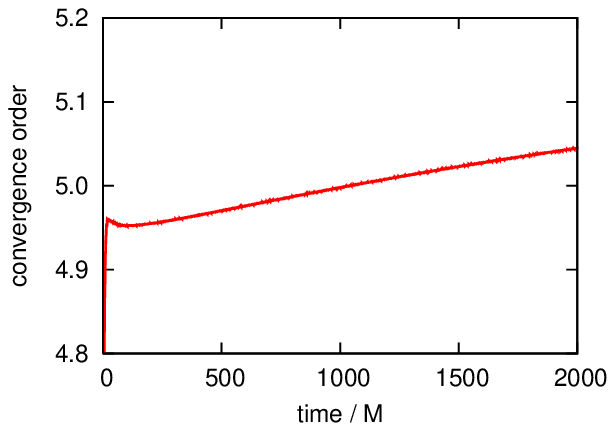}
\caption{$L_2$ norm (top panel) and convergence factor (bottom panel)
  for the Hamiltonian
constraint for evolutions of a Kerr black hole with spin $j=0.5$. 
The coarse resolution corresponds to $16\times 16$ points per block in the angular
directions and $\Delta r = 0.2\,M$ in the radial direction.
The fine resolution a factor of $1.5$ higher in all directions.  We see fifth
order convergence, as expected for the difference operators used.}
\label{fig:convtest-kerr}
\end{center}
\end{figure}

In order to test both the long term stability and the convergence of
our code, we
first evolve a Kerr black hole in Kerr--Schild coordinates with spin
$j=0.5$.  Fig.~\ref{fig:convtest-kerr} shows the $L_2$ norm of the Hamiltonian
constraint vs.\ time for two different resolutions.  The radial domain
extent is
$1.8\,M < r < 11.8\,M$.  The coarse resolution corresponds to $\Delta r
=0.2\, M$ and 
$16 \times 16$ points per block in the angular directions, and the
fine resolution increases the number of points in all directions by a factor
of $1.5$.  We see approximate fifth order convergence, as expected.

In the simulations discussed below, we place our inner boundary at
$r=1.8\,M$ and our outer boundary at $r=251.8\,M$.  This allows for
observer locations up to $r=80\,M$, which are still causally
disconnected from the outer boundary for times long enough to follow
the ringdown, namely up to $t=280\,M$.  We set up initial data
according to eq.~(\ref{eq:InitialData}) with $A=0.01$,
$\sigma=1.0\,M$, and $r_0=20\,M$, where $M$ is the mass of the 
black hole when the perturbation is switched off.  Our coarse resolution uses
$16\times16$ points per block in the angular directions and $1251$
points in the radial direction, corresponding to $\Delta r = 0.2\,M$.
Our fine resolution uses $1.4$ times as many grid points in all
directions.

One of the goals of the analysis that follows is to study the effect
of the choice of the background metric on the accuracy of the
waveforms.  Since for this scattering problem solutions in closed form
are not known, we compare the waves which we extract from our
three-dimensional simulations to results obtained with an independent
fourth order accurate one-dimensional code which solves the
Regge--Wheeler equation (\ref{master}).  These 1D results were
obtained with a resolution of $\Delta r = 0.0125\,M$.  The relative
difference in this Regge--Wheeler function to a result from twice this
resolution lies roughly between roundoff error and $10^{-7}$, which is
far below the numerical errors that we expect from our 3D simulations.
Therefore, we consider these 1D results in the following to be exact
for all practical purposes.

\subsection{The standard and generalized RW  approaches: numerical comparisons}
\label{sec:NumericalComparison}

We now analyze the results of evolving distorted black holes as
described above and extracting gravitational waves with different
methods. 

\begin{figure}
\begin{center}
\includegraphics[width=0.45\textwidth]{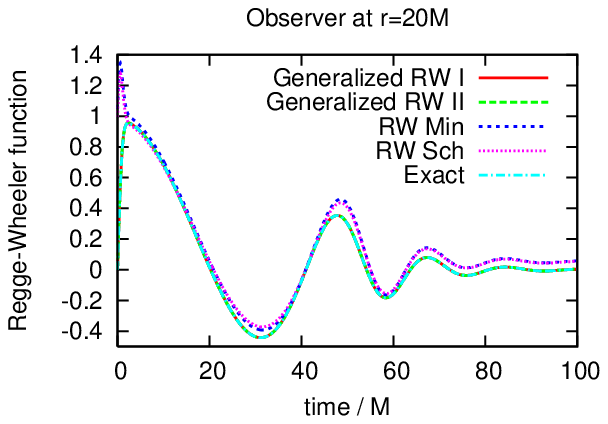}
\includegraphics[width=0.45\textwidth]{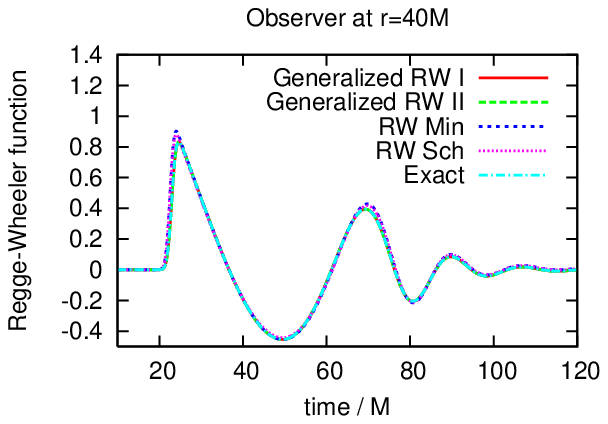}
\includegraphics[width=0.45\textwidth]{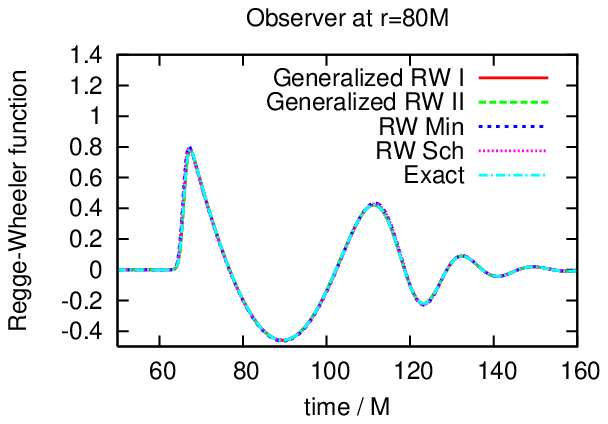}
\caption{Extracted waveforms for observers at $20\,M$, $40\,M$, and
  $80\,M$. Shown is the
Regge--Wheeler function obtained from the standard RW approach and our
generalized one. For the former we assumed both a Minkowski background
and a
Schwarzschild background in Schwarzschild coordinates, labeled as \emph{RW Min} and
\emph{RW Sch}, respectively.  For the generalized approach we show the
results for two cases, in which the background metric is
dynamically computed from
the numerical solution (\emph{Generalized RW I}), and where
we prescribe it analytically 
(\emph{Generalized RW II}). Also
shown is the exact waveform. These simulations were performed
with a resolution of $16\times16$ grid points in the angular
directions on each block and $\Delta r = 0.2\,M$ in the radial
direction.
See the main text for more details.}
\label{fig:waves}
\end{center}
\end{figure}

Fig.~\ref{fig:waves} shows Regge--Wheeler functions for observers
at $r=20\,M$, $40\,M$, and $80\,M$, extracted with both our 
generalized approach and the standard one. The data have been scaled by a
factor of $100$ to normalize to an initial
data amplitude $A=1$ in eq.~(\ref{eq:InitialData}).
Recall that we used weak waves of amplitude 
$A=0.01$ for these simulations to avoid non-linear effects, and
to be able to compare with the exact solution, which is only known in the
linear regime.

Five waves are shown in
fig.~\ref{fig:waves} for each observer location.  Apart
from the exact solution, we show two results obtained
from our generalized approach, which coincide with each other in the
continuum limit.  They differ in how the background
metric is computed: in one case we use the exact expressions for the
Kerr--Schild background, and in the other case these
coefficients were numerically calculated by extracting the $\ell=0$ part
of the metric, as explained in sect.~\ref{sec:extract}. 

Finally, two
waveforms were extracted using the standard
approach with two different assumptions for the background, as found in
the literature: a Minkowski spacetime in Minkowski coordinates, and a
Schwarzschild spacetime in Schwarzschild
coordinates.  We want to highlight an interesting feature which can
easily be seen in eq.~(\ref{pot_adm}).  For any observer location, the waves
extracted with these two background should differ only by a
factor which is constant in time:
\begin{eqnarray}
  \Phi_{RW}^{\Min} & = & \kappa \Phi_{RW}^{\Sch} \,,
\end{eqnarray}
where $\kappa ^2 = g_{rr}^{\Sch}$ is radial component of 
the Schwarzschild metric in Schwarzschild coordinates.  Such a simple
relationship is a direct consequence of the vanishing
radial shift for these backgrounds.  We confirmed this
expected behavior numerically with high accuracy: at all times and 
for all observers we recover this expected ratio between the two waves
to double precision roundoff error.

\begin{figure}
\begin{center}
\includegraphics[width=0.45\textwidth]{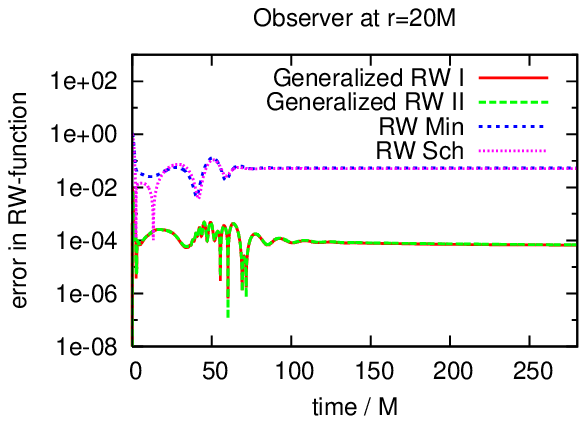}
\includegraphics[width=0.45\textwidth]{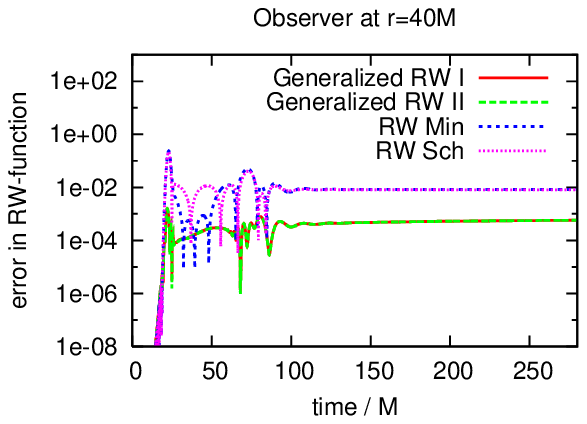}
\includegraphics[width=0.45\textwidth]{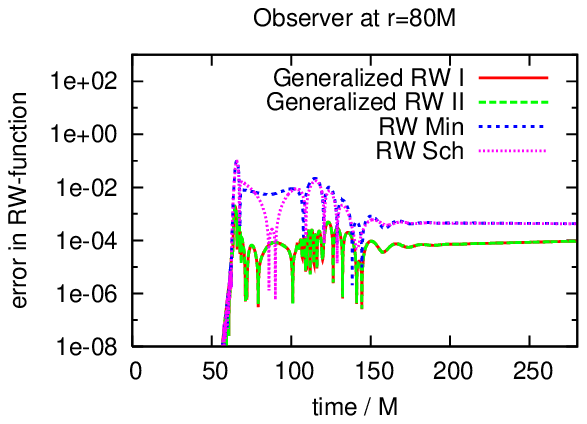}
\caption{Errors for the waveforms shown in
  fig.~\ref{fig:waves}.}
\label{fig:WaveErrors}
\end{center}
\end{figure}

Figure~\ref{fig:waves} suggests that, as expected, 
the differences between
waves extracted with different methods decrease as the
extraction radius increases.  At
$r=80\,M$, the curves show excellent agreement in the 
$L_{e}$ norm\footnote{Also denoted by $L_{\mathrm{\mathbf{e}yeball}}$}.
For a more thorough comparison,
we look at the differences between
the extracted waves and the exact solution in fig.~\ref{fig:WaveErrors}.
For consistency with fig.~\ref{fig:waves}, we also 
scaled the errors relative to the initial amplitude of the perturbation.

Perhaps the most notable feature in fig.~\ref{fig:WaveErrors}  
is that the differences between the waves obtained from generalized approach
either with a numerically obtained background metric or with the exact
(Kerr--Schild) background metric are smaller than the difference
to the exact solution.
For all practical purposes we can therefore consider them  
identical to each other, and for the rest of the paper we leave the
latter out of the discussion. 

Fig.~\ref{fig:WaveErrors} also shows that the
standard approach---with either a Minkowski or Schwarzschild
background---leads to errors which are considerably larger than the errors
in our generalized approach, even for an observer at $r=80\,M$. 
For the specific resolution that we used for fig.~\ref{fig:WaveErrors},  
the errors at $r=20\,M$ with the standard method are roughly three
orders of magnitude
larger than the errors with the generalized method.
For $r=40\,M$ and $80\,M$, the ratio of the errors
is of the order $10^{3}$ to $10^{1}$ and $10^{2}$ to $10^{0}$, respectively.

\begin{figure}
\begin{center}
\includegraphics[width=0.45\textwidth]{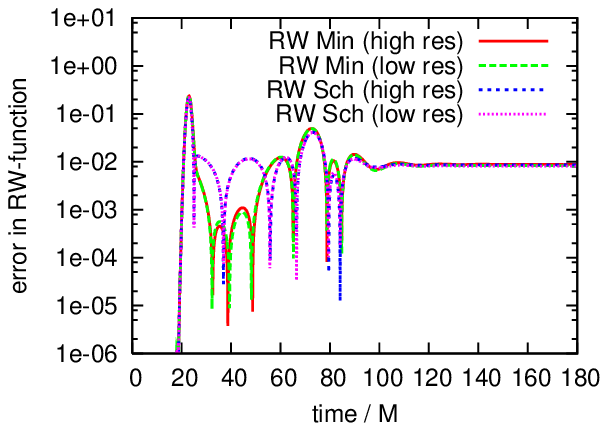}
\includegraphics[width=0.45\textwidth]{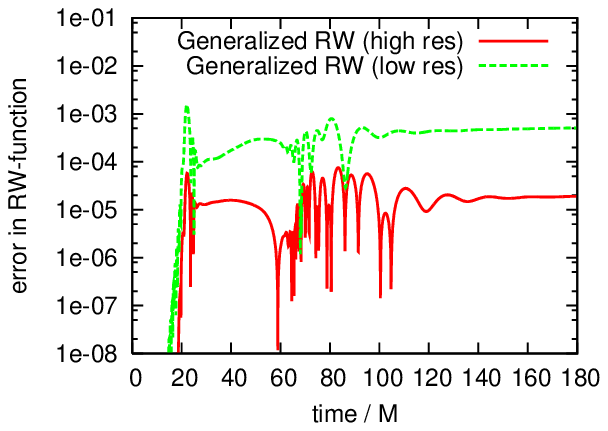}
\caption{Shown is a convergence test for the simulations
  presented in the previous two figures.  The plots labeled with
``low res'' coincide with the ones shown in the previous figures, while 
the plots labeled with ``high res'' correspond to $1.4$ times that
resolution. The error in the generalized wave extraction method, which by design gives the 
correct waveform in the continuum for these simulations, converges towards zero as
expected. On the other hand, the errors 
in the standard wave extraction method are almost unaffected by the increased resolution. This
indicates that these errors are dominated by the extraction method
itself, not by the numerical truncation
error. These results correspond to 
an observer at $40\,M$, but they look similar for the other extraction
radii that we consider in this paper.}
\label{fig:WaveErrorsConvergence}
\end{center}
\end{figure}

The previous discussion only analyzes the errors introduced by the
standard method at a fixed resolution. Next we discuss the dependence
of these results on the resolution. It turns out that the
difference between the different methods is even more striking
for higher resolutions.
By construction, the generalized
wave extraction method should give the exact waveform in the
continuum. At the discrete level, its associated errors should
converge away with increasing resolution.  
Fig.~\ref{fig:WaveErrorsConvergence} shows that this is actually 
the case. On the other hand, the errors in the standard
approach do \emph{not} converge to zero, as shown in 
fig.~\ref{fig:WaveErrorsConvergence}.  In other words, the accuracy
of the extracted waves with the standard method 
is dominated by the extraction procedure and
not by the numerical resolution. 
 
Fig.~\ref{fig:WaveErrorsConvergence} as well as the second 
panel of fig.~\ref{fig:WaveErrors} show another interesting feature. 
Contrary to expectation, assuming Schwarzschild--like coordinates instead of a Minkowski
background does not necessarily lead to smaller errors in the waveforms. 
For example, for an observer at $r=40M$ and during the time interval of  
about $25M<t<50M$, the errors are actually up to one order of magnitude larger
for the  Schwarzschild--like coordinates. However, as can be seen
from fig.~\ref{fig:WaveErrors}, this feature depends on the observer
location.  We assume that this feature is only a coincidence.
 
The plateau in the errors seen in the last $100\,M$ to $200\,M$ in
fig.~\ref{fig:WaveErrors} is due to an offset
in the waveform. We found that, once the wave function decays to 
a small enough amplitude, it no longer oscillates around zero, but instead
oscillates around a certain offset. This can be seen more clearly from the 
top  panel in fig.~\ref{fig:waves}. This offset is present for both the
standard and the generalized extraction methods; however, there are important
differences.  The first is that the offset for the generalized extraction
the offset converges to zero with increasing
resolution, unlike for the standard method.
The other is that the offset for generalized method 
is orders of magnitude smaller than for the standard method.
As we will discuss in the next subsection, that has direct 
consequences when attempting to extract quasinormal frequencies. This offset
is reminiscent of the one that is present in RWZ waveforms when there is
spin \cite{Alcubierre00b, Brandt94c}. 

The oscillatory feature of the wave
can be followed for a longer time if the offset is subtracted from
the waveform  by hand, that is, if the wave is shifted along the vertical axis so that it oscillates
around zero at late times. We do so by fitting the data to an exponentially decaying
wave with an offset.  (Details about the fit are given in the 
following subsection)
The actual values that we determined for the offset
are given in table~\ref{tab:Offset}. As expected, the offset is decreasing
with increasing radius for both standard RW wave extraction methods. 
This offset is mainly a result of the wrong assumption about the
background metric, not of numerical error.
There is no such clear dependence on the radius when using
the new generalized extraction. Here the offset originates 
solely from truncation error, and converges
to zero with increasing resolution. This behavior can also be seen in 
fig.~\ref{fig:WaveErrorsConvergence}.

\begin{table}[htdp]
\caption{Values of the offset for different wave extraction methods and observers
at $20M$, $40M$ and $80M$.}
\begin{center}
\begin{tabular}{|r|r|r|}
\hline
Extraction Method &    Observer &     Offset \\
\hline \hline
Generalized RW &         $20M$ &   $-7.1\times 10^{-5}$ \\

Generalized RW &         $40M$ &    $5.6\times 10^{-4}$ \\

Generalized RW &         $80M$ &    $8.9\times 10^{-5}$ \\
\hline
      RW Min &         $20M$ &   $-5.4\times 10^{-2}$ \\

      RW Min &         $40M$ &   $-8.3\times 10^{-3}$ \\

      RW Min &         $80M$ &   $-4.4\times 10^{-4}$ \\
\hline
     RW Sch &         $20M$ &   $-5.1\times 10^{-2}$ \\

     RW Sch &         $40M$ &   $-8.1\times 10^{-3}$ \\

     RW Sch &         $80M$ &   $-4.3\times 10^{-4}$ \\
\hline
\end{tabular}  
\end{center}
\label{tab:Offset}
\end{table}

\begin{figure}
\begin{center}
\includegraphics[width=0.45\textwidth]{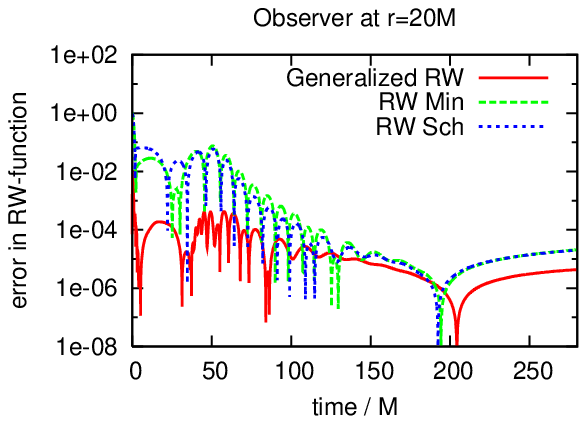}
\includegraphics[width=0.45\textwidth]{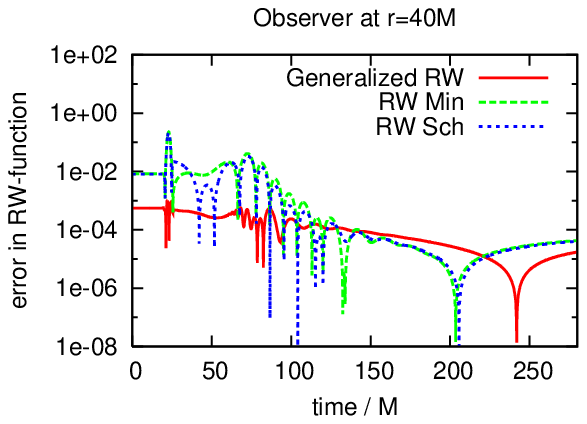}
\includegraphics[width=0.45\textwidth]{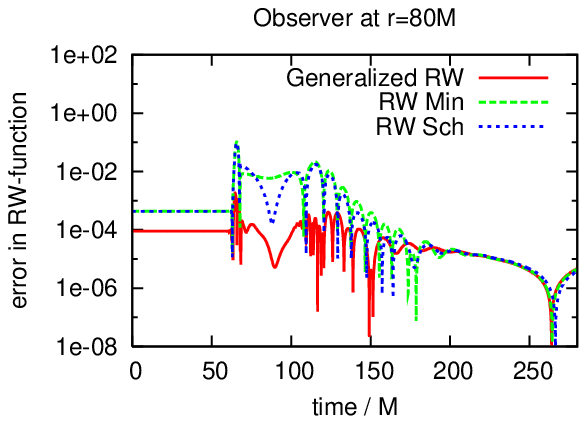}
\caption{Shown are the same quantities as in fig.~\ref{fig:WaveErrors}, 
except that an 
offset is subtracted from each waveform before calculating the
errors. See the main text for details.}
\label{fig:WaveErrorsOffset}
\end{center}
\end{figure}

In fig.~\ref{fig:WaveErrorsOffset} we show the 
difference between the waveforms shifted by different offset values and the 
exact solution, for the same observers as before. As can be seen from
the figure, our
qualitative statements about the accuracies of the different wave extraction
methods remain unchanged, if you consider the time span during which
the amplitude of the wave is significant.%
\footnote{Of course, because we subtract the offset by hand to decrease the
errors at late times, 
we can naturally follow the oscillatory part of the
wave for longer times before.}  We conclude that
the main errors in fig.~\ref{fig:WaveErrors} are \emph{not} caused by an
overall offset in the whole waveform.

\subsection{Quasinormal frequencies}
We now turn our attention to extracting quasinormal
frequencies from the waveforms just discussed. 
 The primary goal is to 
find out whether these frequencies 
are affected by the choice of a specific wave extraction method,
which may have some
presumably small but non-vanishing systematic error for any finite
extraction radius, and if so, by how much. 
We used data from the lower resolution run 
that we already analyzed in the
previous section. The accuracy of the frequency does not change significantly 
if we use the higher resolution run instead. 

The angular part of the initial data is a pure $\ell=2$, $m=0$  
mode. Since the background has no angular momentum, there is no
mode--mode coupling
at the linear level, while nonlinear coupling
can be neglected for the current study, because we only evolve weak
perturbations.
Therefore the only dominant multipole
mode present in the data at all times should 
be the one injected initially. At the numerical level, $\ell=4$ modes
can be generated by our six-block grid structure. However, in
\cite{Dorband05a} it was found 
that in the absence of angular momentum, these modes not only
converge to zero with
resolution, but are also very small for the resolutions
considered in this paper. In the above reference and in \cite{Berti:2006wq}
it was also shown that overtones are not
significantly excited unless the black hole is very rapidly
rotating. Based on all this, we only fit for a single   
$\ell=2$, $m=0$ mode:
\begin{equation}
\Psi_{RW}^{fit} = \mathcal{A} \sin(\omega_r t + \chi)\, e^{\ii \omega_i (t - t_0)} - \xi
\label{eq:fit}
\end{equation}
where $\mathcal{A}$ is the excitation amplitude, 
$\omega = \omega_r + \ii \omega_i$ is the complex quasinormal mode
frequency, $\chi$ is a phase shift, $\xi$ is the offset and $t_0$ is the starting time 
of the quasinormal ringing regime. The latter is not unambigously defined
 (the so called ``time-shift problem''), and as a
consequence neither are the amplitudes of quasinormal modes. 
In ref.~\cite{Dorband05a} it was proposed to minimize the uncertainties due to
this time-shift problem by looking at carefully chosen relative
amplitudes (see that reference for details).
In order to fit numerical data to eq.~(\ref{eq:fit}), we fix $t_0$ to
an educated
guess\footnote{For example, taking into account where the
  initial data and observer are located, and assuming a propagation speed of
  one} and then fit
for $\omega$, $\mathcal{A}$, $\chi$, and $\xi$. Any difference in $t_0$
is absorbed in $\mathcal{A}$ (in which we are not interested at this point)
and does not change the other extracted parameters. 
We find the time-window of optimal
fitting by looking for a local minimum in the relative residual
between the original waveform and its fit. In ref.~\cite{Dorband05a}
it was found that such a local minimum is usually quite sharp and
therefore gives a
good criteria for choosing the window of time where the quasinormal
ringing dominates. Similarly, we use the uncertainties in this minimum to
quantify the errors in the parameters obtained in the fit. More
details about the fitting procedure that we use to extract quasinormal
parameters are given in ref.~\cite{Dorband05a}. 

In the previous subsection we discussed the presence of an offset in the extracted waves
with the standard method. If such an offset is not taken into account
when fitting for the quasinormal frequencies (i.e., for a fixed $\xi = 0$), 
eq.~(\ref{eq:fit}) does not represent the behavior of the numerical data well
enough, and no reasonable results can be obtained from the fit. This is
 especially the case 
at medium to late time intervals when the amplitude becomes smaller than the
offset, so that the wave does not cross zero any more. When one tries to 
fit for these cases, the obtained frequency has no relation at all
to the 
correct QNM frequency. For example, at $r=20M$ the offset in the  
waves obtained from the standard RW wave extraction is of 
order $10^{-2}$ for both a Minkowski background and for Schwarzschild--like
coordinates. Without taking the offset into account, the value of 
$\omega_r$ that the fit determines lies between
$10^{-14}$ and $10^{-4}$, and $\omega_i$ is of order $10^{-3}$ to
$10^{-6}$ (compare to table \ref{tab:QNM}). 
In contrast, the offset resulting from the generalized RW wave
extraction is of 
order $10^{-5}$
for this resolution. This is small enough that the problems
described above do not play a noticeable role.

\begin{table*}[htdp]
\caption{Quasinormal frequencies of the $\ell=2, m=0$ mode as measured by an observer at
$r=20\,M$. Results are given for waveforms resulting from the different extraction methods 
we use.
The predicted frequency from perturbation theory, which we assume to be
exact because our perturbation amplitude is small, is 
$\omega_{\mathrm{exact}} = 0.37367-0.08896\ii$ \cite{Leaver85}. The
uncertainties in the extracted frequencies originate from
variations in them depending on which interval of the waveform is used
for the fit. The relative error is defined as
$\left| (\omega-\omega_{\mathrm{exact}})/\omega_{\mathrm{exact}}
\right |$.
}
\begin{center}

\begin{tabular}{|l|l|l|l|}
\hline
Extraction Method             & $\omega$ & relative error \\\hline\hline
Generalized RW        & $0.3736-0.0890\ii \pm (3+3\ii)\times10^{-4}$ & $1.9\times10^{-4} + 4.5\times10^{-4} \ii$ \\
RW Min       & $0.3733-0.0889\ii \pm (3+3\ii)\times10^{-4}$  & $9.9\times10^{-4} + 6.7\times10^{-4} \ii $\\
RW Sch       & $0.3733-0.0889\ii \pm (3+3\ii)\times10^{-4}$  & $9.9\times10^{-4} + 6.7\times10^{-4} \ii $\\
\hline
\end{tabular}  

\end{center}
\label{tab:QNM}
\end{table*}

Table \ref{tab:QNM} shows the complex quasinormal frequencies that we obtained from the 
generalized and from the standard RW methods. As mentioned above and discussed
in detail in ref.~\cite{Dorband05a}, the 
error bars are estimated from changes in the frequency when changing
the time interval that we use for the fit of the waveform.
We assume that the predicted frequency from perturbation theory for
the fundamental $\ell=2$, $m=0$ mode is exact because we use a small
amplitude for our perturbation.  This frequency is known to be
$\omega_{\mathrm{exact}}=0.37367-0.08896\ii$ (see for example  \cite{Leaver85}).
The frequency obtained from the new generalized wave extraction is consistent with 
this exact value within the accuracy to which we can obtain these
numbers from the fit itself. For the 
standard wave extraction method, we only find agreement to three
significant digits in the real part, but
better agreement with the exact value in the 
imaginary part of the waveform. Note that, since the waveforms only differ
by a constant factor  (see subsec.~\ref{sec:NumericalComparison}), the frequencies obtained  with
 a Minkowski and a Schwarzschild
background agree to roundoff error. 
The reason for the lower accuracy in the
real part of $\omega$ might be due to the fact that the waveforms
are slightly distorted due to the wrong assumption for the background metric. This causes
a larger residual between the data and the fit---it is about a factor of two larger 
than with the generalized wave extraction---and some degradation in how accurately
certain fitting unknowns like $\omega$ can be determined. That may also explain the 
larger relative error for the waves extracted with the standard RW wave method, 
which is shown in the right column of the
same table. There the relative error is 
defined as $\left| (\omega-\omega_{\mathrm{exact}})/\omega_{\mathrm{exact}} \right |$.

\section{Final comments}
\label{sec:conclusions}

When considering methods for extracting gravitational waves from
numerical spacetimes at a finite distance, one question of direct
interest is: How sensitive is the accuracy of the extracted waveforms
to both the extraction method and observer location?  In particular,
how far away is ``far enough'' when extracting gravitational waves?

It is in general not easy to pose such a question in a precise way,
since in order to quantify this one needs an exact waveform to compare
with.  This exact waveform is in principle only well defined at future
null infinity.  However, there are some particular scenarios of
interest where the concept of ``exact waveforms'' at a finite distance
can be given a well defined and precise sense.  That is the case, for
example, for perturbations of Kerr black holes (actually, of Petrov
type $D$ spacetimes): the Weyl scalar $\Psi_4$ is defined everywhere
in an essentially gauge and tetrad invariant way \cite{Chandrasekhar83}.  
Similarly, for
perturbations of Schwarzschild black holes, the Regge--Wheeler and
Zerilli functions are defined in a gauge invariant way everywhere as
well.  In fact, there is a one-to-one mapping between these functions
and $\Psi_4$; see e.g.\ \cite{Sarbach:2001qq}.

Therefore the above question can be posed in a setting that might
not be the most general one, but it is one in which a precise,
quantitative answer can be found.  The concrete setting that
we chose for the current study is that of weak 
perturbations of Schwarzschild black holes.  Furthermore,
in this paper we restricted our treatment to odd parity
perturbations (the even parity sector will be presented elsewhere).  
One of the standard methods that has been widely
used for extracting gravitational waves from such spacetimes is through
the standard Regge--Wheeler--Zerilli perturbation formalism.  This formalism
provides a gauge invariant treatment of perturbations of a background
geometry defined by the Schwarzschild spacetime \emph{in Schwarzschild
coordinates}.  That is, the formalism is invariant with respect to
linear coordinate transformations \emph{which leave the background
  coordinates fixed}.  If one extracts waves using this
formalism from a perturbation of Schwarzschild in, say, Kerr--Schild
coordinates, the extracted waves at a finite distance will not be
correct, even with if extracted with infinite numerical
precision.  There is a systematic error in such an extraction, due to the
incorrect identification of the background coordinates.  Of course, one
expects this error to
decrease as the extraction radius increases.  In the spirit of the
above discussion, the question that we asked ourselves was: how
far away must the observer be, so that the difference between the exact waveform
and the extracted one is negligible, if the extraction method has
a systematic error?  For this study we chose a very specific 
interpretation of ``negligible
systematic errors in the waveforms'', namely, that they are smaller
than or comparable to the errors in these waveforms
due to the numerical discretization. 

In order to provide a quantitative answer to this question, in this
paper we first proceeded to generalize the standard Regge--Wheeler
extraction approach by using a perturbation formalism that allows  
for quite general slicing conditions for
the Schwarzschild background.  With this generalization, if one calculated
with infinite resolution, one would extract the exact waveforms for
\emph{any} (not necessarily large) finite extraction radius.
This holds even if the Schwarzschild background is, for example, given
in a time
dependent slicing, or one in which the coordinates in a neighborhood of
the horizon are well defined, as is usually the case in numerical
black hole evolutions. 

After summarizing the basics of the generalized formalism, we
described our numerical implementation of the generalized
extraction mechanism and our way of solving Einstein's
equations.  For the latter we used multiple blocks and high order
methods, both of which present several advantages.  Of particular
interest to this paper is that, due to the
adaptivity that multiple blocks provide, the outer boundary can be
placed at large distances, with much smaller computational costs
than with Cartesian grids and mesh refinement.  We made use of this
specific advantage and performed three-dimensional non-linear simulations of
weakly distorted Schwarzschild black holes, from which we extracted waves at
distances larger than
most current state of-the-art three-dimensional simulations of
Einstein's equations.

Then we studied the dependence of the extracted
waveforms on the extraction method.  More precisely, we compared the
standard RW method with our generalized one.  We found that, even for the
coarsest resolutions that we used, the errors in the waveforms from the
standard method were dominated by the extraction procedure and
not by the numerical accuracy of the spacetime metric.  Furthermore, by
increasing the resolution we could explicitly demonstrate that the
errors in the standard
method do not approach  zero, while they do with the
generalized one.
While this is obviously the expected behavior on
analytical grounds, we emphasize that we could explicitly see
these differences even with an extraction radius which is significantly larger
than those typically used in current state-of-the-art
three-dimensional simulations. 

What is not clear, however, is whether the wave zone resolution currently used
by mesh refinement codes is sufficient to see the differences that we
have demonstrated in this paper.  For example, the spatial
resolutions in the wave  zone of current binary black hole
inspiral and coalescence simulations are usually much coarser than the
resolutions that we used above.  Some radial resolutions $h$ in the
wave zone of binary black hole system simulations are:
\cite{Baker05a} $h=0.5\,M$,
\cite{Herrmann:2006ks} $h=0.5\,M$ (but the extraction is performed
very close in at $R=16M$)
\cite{Baker:2006yw} $h=0.75\,M$ (but $h=1.5\,M$ for calculating the
radiated angular momentum $J$),
\cite{Pretorius:2006tp} $h=0.85\,M$,
\cite{Sperhake:2006cy} $h=0.87\,M$
\cite{Buonanno06imr} $h=0.82\,M$,
\cite{Bruegmann:2006at} $h=0.56\,M$,
\cite{Gonzales06tr} $h=0.56\,M$.
Some of these codes are 4th order accurate, but many have at least
certain components that are only 2nd order accurate.%
\footnote{While it is currently common practice to report the finest
  resolution (near the horizons) and the coarsest resolution (near the
  outer boundary) in such simulations, the resolution in the wave
  zone, i.e., at the location where the gravitational waves are
  extracted, is often not explicitly listed, and can sometimes not be
  inferred.  Some publications also do not report at which coordinate
  radius the wave information is extracted.}

One of the interesting features of the waveforms that we extracted in
this paper with the standard method is that we were able to
``postprocess'' them in order to remove an offset at late times.
By doing so, we could accurately extract the quasinormal
frequencies.  However, we explicitly demonstrated that
the large errors in the standard method were not due to an overall offset in
the whole wave.  Even after removing the offset ``by hand'',
errors of roughly the same order in the waves remained at early and
intermediate times in the ringing regime.  In addition, this
postprocessing made use of the fact that we knew the qualitative
behavior of the exact solution in the quasinormal ringing regime.  In
particular, we knew that it had to oscillate around zero, and we also knew
what the frequencies they were supposed to have.  It is not clear that one
could apply
such a postprocessing to decrease systematic errors in a more general
scenario, where the characteristics of the expected waveforms are
either completely unknown or not
known with so much detail.

Concluding,
in this paper we considered weak perturbations of Schwarzschild black holes,
for which---as mentioned above---one can construct the Regge--Wheeler and
Zerilli functions (or, equivalently, $\Psi_4$) in an unambiguous way {\em
everywhere}. In a more generic case (say, a collision of compact
objects) this is not possible, and all gravitational wave
extraction methods are  inherently approximate at a fixed finite distance. 
The results of this paper suggest that, depending on the accuracy
of a given simulation, different choices in the extraction procedure
at a fixed and finite distance may result in relative differences in the
waveforms that are actually larger than the numerical errors of
the solution. These differences will in general decay with radius, but in a
very slow way; typically as $1/r$ (which is, in fact, the decay
we found in our simulations). For example, in order to decrease the
systematic errors for an observer at $40\,M$ shown in 
fig.~\ref{fig:WaveErrorsConvergence}  by,
say, two orders of magnitude, by just moving the observer out and
extracting at a single extraction radius, the
latter would have to be located at $\simeq 4,000\,M$. 
This means that, if similar uncertainties show up in other simulations
for differing extraction methods, as the results of this paper
suggest (and which can be tested), then decreasing those uncertainties
by extracting waves at a {\em fixed} location and moving 
 the observer further out does not seem feasible, and other ideas
would have to be explored.

\vspace*{3ex} 
\begin{acknowledgments}

  We thank Emanuele Berti, Steve Brandt, Barrett Deris, 
  Luis Lehner, Jorge Pullin, Olivier Sarbach, and Saul Teukolsky for many helpful
  discussions and suggestions. 
  
  This research was supported in part by NSF grants PHY-0505761, PHY-0244699, 
  PHY-0326311, PHY-0554793 
  and the National Center for Supercomputer
  Applications grant MCA02N014 to Louisiana State University.  It also
  employed the resources of the Center for Computation \& Technology
  at Louisiana State University, which is supported by funding from
  the Louisiana legislature's Information Technology Initiative. AN thanks the
  CCT for an extended visit at Louisiana State University, where part
  of this work was done.  MT
  thanks Saul Teukolsky for hospitality at
  Cornell University, and Reinaldo
  Gleiser and Oscar Reula for hospitality at FaMAF, where parts of this
  work were done. 
  Our numerical calculations used the Cactus framework
  \cite{Goodale02a,cactusweb1} with a number of locally developed
  thorns, the LAPACK \cite{lapackweb} and BLAS \cite{blasweb}
  libraries from the Netlib Repository \cite{netlibweb}, and the LAM
  \cite{burns94:_lam, squyres03:_compon_archit_lam_mpi, lamweb} and
  MPICH \cite{Gropp:1996:HPI, mpich-user, mpichweb} MPI \cite{mpiweb}
  implementations.

\end{acknowledgments}

\clearpage

\appendix

\begin{widetext}
\section{Vector and tensor spherical harmonic decomposition (odd
  \emph{and} even-parity sectors)} 
\label{harmonic_decomposition}
We discuss now, in some detail, how to compute a multipole decomposition using vector and
tensor spherical harmonics. A vector field $V_A$ defined on the manifold
$S^2$ can be decomposed in multipoles using even and odd-parity basis vectors. Denoting
the components in this basis by $h^{\lm}_{\even}$ and $h^{\lm}_{\odd}$, $V_A$ 
can be written as
\begin{equation}
V_A = \sum_{\ell=1}^{\infty}\sum_{m=-\ell}^\ell h^{\lm}_{\even} Y_A^{\lm} + h^{\lm}_{\odd} S_A^{\lm}.
\end{equation}
Here, $Y_A^{\lm}$ and $S_A^{\lm}$ are the even and odd-parity basis vectors tangent 
to the sphere, respectively. They are defined as
\begin{eqnarray}
Y_A^{\lm} &=& \nabhat_A Y^{\lm} \\
S_A^{\lm} &=& \epshat^B_{\; A} \nabhat_B Y^{\lm},
\end{eqnarray}
where $\nabhat_A$ is the covariant derivative compatible with the unit sphere metric $\ghat_{AB}$
and  $\epshat_{AB}$ is the Levi-Civita tensor with components $\epshat_{\vt \vp} = \sin \vt$.
They satisfy the relations $\nabhat_C \ghat_{AB}=0$ and $\nabhat_C \epshat_{AB}=0$. These
vectors obey the orthogonality relations
\begin{eqnarray}
\int \ghat^{AB} \bar{Y}^{\lm}_A Y^{(\ell' m')}_B d\Omega &=& \ell(\ell +1) \delta_{\ell \ell'}
  \delta_{m m'}, \\
\int \ghat^{AB} \bar{S}^{\lm}_A S^{(\ell' m')}_B d\Omega &=& \ell(\ell +1) \delta_{\ell \ell'}
  \delta_{m m'}, \\
\int \ghat^{AB} Y^{\lm}_A S^{(\ell' m')}_B d\Omega &=& 0.
\end{eqnarray}
Here $d\Omega=\sin \vt \, d\vt \, d\vp$ is the area element in polar spherical
coordinates and the bar denotes complex conjugation. Using the orthogonality 
property we can find the multipole modes $h^{\lm}_{\even}$ and
$h^{\lm}_{\odd}$. The result is
\begin{eqnarray}
h^{\lm}_{\even} &=& \frac{1}{\ell(\ell +1)} \int \ghat^{AB} V_A \bar{Y}_B^{\lm} \, d\Omega, \\
h^{\lm}_{\odd}  &=& \frac{1}{\ell(\ell +1)} \int \ghat^{AB} V_A \bar{S}_B^{\lm} \, d\Omega.
\end{eqnarray}
Using spherical coordinates the components of the even-parity basis are
\begin{eqnarray}
Y^{\lm}_\vt &=& \partial_\vt Y^{\lm} \\
Y^{\lm}_\vp &=& \partial_\vp Y^{\lm}.
\end{eqnarray}
For the odd parity we get
\begin{eqnarray}
S^{\lm}_\vt &=& -\frac{1}{\sin \vt} \partial_\vp Y^{\lm} \\
S^{\lm}_\vp &=&  \sin \vt \; \partial_\vt  Y^{\lm}.
\end{eqnarray}
Expanding the integral with these vector components we obtain that
\begin{eqnarray}
h^{\lm}_{\even} &=& \frac{1}{\ell(\ell +1)} \int V_\vt
\bar{Y}_\vt^{\lm} + \frac{1}{\sin^2\vt}V_\vp \bar{Y}_\phi^{\lm} \, d\Omega, \\
h^{\lm}_{\odd}  &=& \frac{1}{\ell(\ell +1)} \int \frac{1}{\sin\vt}\left(V_\vp 
  \bar{Y}_\vt^{\lm} - V_\vt \bar{Y}_\vp^{\lm}\right) \, d\Omega.
\end{eqnarray}
For tensors, the idea is the same. If $V_{AB}$ is a tensor field defined on the unit 
sphere, the multipole decomposition takes the form
\begin{equation}
V_{AB} = \sum_{\ell=2}^\infty \sum_{m=-\ell}^{\ell} K^{\lm} \ghat_{AB} Y^{\lm} 
  + G^{\lm} Y^{\lm}_{AB} + h_2^{\lm} S^{\lm}_{AB},
\end{equation}
where $\ghat_{AB} Y^{\lm}$ and $Y^{\lm}_{AB}$ are the even-parity tensor basis,
whereas $S^{\lm}_{AB}$ is the odd-parity tensor basis. We follow the 
Regge--Wheeler notation by
using  $K$ and $G$ for the even-parity components and $h_2$ for the odd-parity one.
The tensor basis is defined as
\begin{eqnarray}
Y^{\lm}_{AB} &=& \nabhat_A \nabhat_B Y^{\lm} + \frac{1}{2}\ell(\ell+1) \ghat_{AB} 
  Y^{\lm} \\
S^{\lm}_{AB} &=& \frac{1}{2}\left( \nabhat_A S^{\lm}_B + \nabhat_B S^{\lm}_A \right)
\end{eqnarray}
This definition agrees with Zerilli tensor harmonics up to a factor of 2, as we will see.
 They obey the orthogonality relations
\begin{eqnarray}
\int \ghat^{AC} \ghat^{BD}\bar{Y}^{\lm}_{CD} Y_{AB}^{\ell' m'} d\Omega &=& \frac{1}{2}
  \ell(\ell -1)(\ell +1)(\ell +2) \delta_{\ell \ell'} \delta_{m m'}, \\
\int \ghat^{AC} \ghat^{BD}\bar{S}^{\lm}_{CD} S_{AB}^{\ell' m'} d\Omega &=& \frac{1}{2}
  \ell(\ell -1)(\ell +1)(\ell +2) \delta_{\ell \ell'} \delta_{m m'},
\end{eqnarray}
and integration of the product of two different tensor basis vanishes. With this we
can find $K$, $G$ and $h_2$. The result is
\begin{eqnarray}
K^{\lm} &=& \frac{1}{2} \int V_{AB} g^{AB} \bar{Y}^{\ell m} \, d\Omega \\
G^{\lm} &=& \frac{2}{\ell(\ell -1)(\ell+1)(\ell+2)} \int  
  \ghat^{AC} \ghat^{BD} V_{AB} \bar{Y}_{CD}^{\lm} \, d\Omega \\
h_2^{\lm}   &=& \frac{2}{\ell(\ell -1)(\ell+1)(\ell+2)} \int  
  \ghat^{AC} \ghat^{BD} V_{AB} \bar{S}_{CD}^{\lm} \, d\Omega
\end{eqnarray}
Using spherical coordinates the components of the  basis are
\begin{eqnarray}
Y^{\lm}_{\vt \vt} &=& \frac{1}{2} W^{\lm} \\
Y^{\lm}_{\vt \vp} &=& \frac{1}{2} X^{\lm} \\
Y^{\lm}_{\vp \vp} &=&-\frac{1}{2} \sin^2 \vt \, W^{\lm} \\
S^{\lm}_{\vt \vt} &=&-\frac{1}{2  \sin   \vt}\, X^{\lm} \\
S^{\lm}_{\vt \vp} &=& \frac{1}{2} \sin   \vt \, W^{\lm} \\
S^{\lm}_{\vp \vp} &=& \frac{1}{2} \sin   \vt \, X^{\lm},
\end{eqnarray}
where $W^{\lm}$ and $X^{\lm}$ are defined by Zerilli \cite{Zerilli70b} as
\begin{eqnarray}
W^{\lm} &=& 2\left[\partial^2_\vt + \frac{1}{2}\ell(\ell+1)\right] Y^{\lm} \\
X^{\lm} &=& 2\partial_\vp \left( \partial_\vt - \cot\vt \right) Y^{\lm}.
\end{eqnarray}
Assuming that $V_{AB}$ is a symmetric tensor and abbreviating the normalization constant
as $L=\ell(\ell -1)(\ell+1)(\ell+2)$, we  expand the integrals to get
\begin{eqnarray}
K^{\lm} &=& \frac{1}{2} \int \left( V_{\theta\theta} +
  \frac{V_{\phi\phi}}{\sin^2\theta} \right)  \bar{Y}^{\lm} \, d\Omega 
  \label{eq:kappa} \\
G^{\lm} &=& \frac{1}{L} \int  
  V_{\theta\theta} \bar{W}^{\lm}  + \frac{1}{\sin^2\theta} \left(
  2 V_{\theta\phi}   \bar{X}^{\lm}  -
  V_{\phi\phi}     \bar{W}^{\lm} \right) \, d\Omega \\
h_2^{\lm}   &=& \frac{1}{L} \int 
  \frac{V_{\phi\phi}}{\sin^3\theta} \bar{X}^{\lm} + 
  2\frac{V_{\theta\phi}}{\sin\theta} \bar{W}^{\lm} -
  \frac{V_{\theta\theta}}{\sin\theta}\bar{X}^{\lm} d\Omega
  \label{eq:eta}
\end{eqnarray}
The $Y^{\ell m}$ are normalized with respect to the
standard metric $\ghat_{AB}$ on $S^2$, an exception being the cases
$\ell=0$ and $\ell=1$; where we choose the normalization such that
$Y^{0,0}=1$, and $\int_{S^2}Y^{1,m} \bar{Y}^{1,m} d\Omega = 4\pi/3$.
\end{widetext}


\bibliographystyle{bibtex/apsrev}


\bibliography{bibtex/references}

\end{document}